\documentclass{sig-alternate-05-2015}
\usepackage{cite}
\usepackage[
  pass,% keep layout unchanged
]{geometry}
\newfont{\mycrnotice}{ptmr8t at 7pt}
\newfont{\myconfname}{ptmri8t at 7pt}
\let\crnotice\mycrnotice%
\let\confname\myconfname%

\usepackage{epsfig,balance}
\usepackage[ruled,lined]{algorithm2e}

\begin{document}
%\conferenceinfo{DIVANet'14,} {September 21--26, 2014, QC, Montreal, Canada.}
%\CopyrightYear{2014}
%\crdata{978-1-4503-3028-2/14/09\...\$15.00.\\
 %http://dx.doi.org/10.1145/2656346.2656353}
%\clubpenalty=10000
%\widowpenalty = 10000
\title{Trust-based Secure Routing in Software-defined Vehicular Ad Hoc Networks}

\numberofauthors{2} %  in this sample file, there are a *total*
% of EIGHT authors. SIX appear on the 'first-page' (for formatting
% reasons) and the remaining two appear in the \additionalauthors section.
%
\author{
% You can go ahead and credit any number of authors here,
% e.g. one 'row of three' or two rows (consisting of one row of three
% and a second row of one, two or three).
%
% The command \alignauthor (no curly braces needed) should
% precede each author name, affiliation/snail-mail address and
% e-mail address. Additionally, tag each line of
% affiliation/address with \affaddr, and tag the
% e-mail address with \email.
%
% 1st. author
\alignauthor
Dajun Zhang, F. Richard Yu, Zhexiong Wei\\
       \affaddr{Depart. of Systems Computer Eng.}\\
       \affaddr{Carleton University, Ottawa, ON, Canada}\\
       %\affaddr{}\\
       \email{dajunzhang@cmail.carleton.ca richard.yu@carleton.ca zhexiong\_wei@sce.carleton.ca}
\alignauthor
Azzedine Boukerche\\
       \affaddr{Sch. of Elect. Eng. Computer Science} \\
       \affaddr{University of Ottawa, Ottawa, ON, Canada}\\
       %\affaddr{Ottawa, ON, Canada}\\
       \email{boukerch@site.uottawa.ca}
%\alignauthor
%       \affaddr{San Antonio, Texas 78229}\\
%       \email{cpalmer@prl.com}
}
% There's nothing stopping you putting the seventh, eighth, etc.
% author on the opening page (as the 'third row') but we ask,
% for aesthetic reasons that you place these 'additional authors'
% in the \additional authors block, viz.
%\additionalauthors{Additional authors: John Smith (The Th{\o}rv{\"a}ld Group, email: {\texttt{jsmith@affiliation.org}}) and Julius P.~Kumquat(The Kumquat Consortium, email: {\texttt{jpkumquat@consortium.net}}).}
%\date{30 July 1999}
% Just remember to make sure that the TOTAL number of authors
% is the number that will appear on the first page PLUS the
% number that will appear in the \additionalauthors section.

\maketitle

\begin{abstract}
With the rising interest of expedient, safe, and high-efficient transportation, vehicular ad hoc networks (VANETs) have turned into a critical technology in smart transportation systems. Because of the high mobility of nodes,  VANETs are vulnerable to security attacks. In this paper, we propose a novel framework of software-defined VANETs with trust management. Specifically, we separate the forwarding plane in VANETs from the control plane, which is responsible for the control functionality, such as routing protocols and trust management in VANETs. Using the on-demand distance vector routing (TAODV) protocol as an example, we present a routing protocol named software-defined trust based ad hoc on-demand distance vector routing (SD-TAODV). Simulation results are presented to show the effectiveness of the proposed software-defined VANETs with trust management.
\end{abstract}

% A category with the (minimum) three required fields
%\category{H.4}{Information Systems Applications}{Miscellaneous}
%A category including the fourth, optional field follows...
%\category{D.2.8}{Software Engineering}{Metrics}[complexity measures, performance measures]

%\terms{Design}
\keywords{
Vehicular ad hoc networks; software-defined networking; security; trust management}

\section{Introduction}
In recent years, with the miniaturization of mobile end devices, mobile ad hoc networks (MANETs) become popular in a wide range of fields. For example, they can be used in military, catastrophes, expedition and so on. A vehicular ad hoc network (VANET) is a type of MANETs in the vehicular environment. With the rising demand of convenient, safe, and efficient transportation, VANETs act as an vital role in intelligent transportation systems \cite{Silva13,Zhuang13,Nidhal13,GDS13}.  Vehicle-to-\noindent infrastructure (V2I) and vehicle-to-vehicle (V2V) are two main communication ways in VANETs. MANET routing protocols, such as ad hoc on-demand distance vector routing (AODV), can also be used in VANETs.

Quality of service (QoS) and security issues are two main challenges in wireless mobile networks \cite{MYL04, YL01, LYH10, YK07, XYJL12, ATV12, LY15, WYS10, GYJ10, BYC12, XYJ12, YTH09, YHT10, LYJ10, YZX11, LYL09, ZYN12_JSAC}. Particularly, network topologies of VANETs always change due to the high mobility of nodes. Meanwhile, VANETs are easy to be attacked by DoS, black-hole, and other attacks \cite{Tyagi2014proceeding}. So mitigating these attacks is necessary to improve the security of VANETs.

Researchers have proposed many security mechanisms in order to enhance the security of VANETs \cite{Wei13,Richard13,Wang13,Boukerche13,HKH10,RPB11,SKM12,NTH13}. The authors of \cite{Wei13} propose an distributed cooperative spectrum sensing scheme, in which the scheme aims to solve the security issues of CR-VANETs. An trust based framework is proposed in \cite{Richard13} that provides a second protection to improve security and maintain privacy of VANETs. Wang \emph{et al.} \cite{Wang13} introduce a field game model to solve the security problems in VANETs. Zheng \cite{Boukerche13} \emph{et al.} present a game theoretic approach to quantitatively analyze the attacking strategies of ad hoc networks.\par

Although many researchers have already done some excellent works on trust-based security schemes in VANETs, they are still hard to ensure safety because most existing security works couple data forwarding with control (e.g., routing protocols and trust management). Recently, software-defined networking (SDN) and virtualization \cite{Kreutz2015proceeding,YYG15,LY15,LYZ15,LY15m,CYL15,CYY15} has become a emerging technology, which enables researchers to solve the above problems. Decoupling the control plane from the forwarding plane is the core idea of SDN, which makes the forwarding plane directly programmable \cite{Xia2015ieeecst}. Since SDN separates the control plane from the forwarding plane, the network nodes only act as efficient forwarding devices \cite{Kreutz2015proceeding}. SDN provides a cost-effective networking approach that aims to reduce the cost of wired and wireless networks and improve the network performance.

In this paper, with the recent advances in SDN, we present a novel framework of  software-defined VANETs with trust management. Specifically, we separate the forwarding plane in VANETs from the control plane, which is responsible for the control functionality, such as routing protocols and trust management in VANETs. As AODV protocol \cite{Perkins2003framework} is frequently used in VANETs, we utilize AODV as an example to execute our proposed SDN-based framework in VANETs. In addition, we  move the AODV control logic and the trust management into the control node. Simulation results demonstrate  that our software-defined trust based on-demand distance vector routing (SD-TAODV) can improve the network performance significantly.\par

The rest of paper is organized as follows: The background information of AODV protocol and SDN are presented in Section 2. Section 3 describes our proposed scheme TAODV, and the combination method of SDN and TAODV is depicted in Section 4. Performance of SD-TAODV is evaluated and compared with the traditional AODV in Section 5. Finally, Some conclusions are given in Section 6. \par

\section{Background}
\label{sect:related_work}
VANETs have self-organization features, without relying on the inherent communication network infrastructures. Meanwhile, VANETs can quickly form networks and build network communications. The MAC and routing protocols are two important components of VANET protocols. The MAC protocols include multiple access with collision avoidance (MACA) \cite{Toulgoat13}, carrier sense multiple access (CSMA) and so on. VANET routing protocols can be divided into two main groups: topology based routing protocols and geographic routing protocols \cite{Paul2011proceeding}. In this section, we use AODV as an example to introduce the VANET routing protocols. In addition, we also describe the basic features of SDN and OpenFlow protocol.
\subsection{Overview of AODV Protocol}
AODV is one of the most frequently utilized routing protocols in VANETs \cite{Perkins2003framework}. The main difference between AODV and other VANET routing protocols is that AODV introduces ``sequence number" concept, which is utilized to avoid the \emph{count to infinity problem} and to prevent rooting loop \cite{Gulliver2005proceeding}. Specifically, each node in AODV must maintain its own routing table that includes routing information about its neighbor nodes. The operating procedure of AODV can be divided into two main operations: route discovery and route maintenance \cite{Cao2010proceeding}.\par

The source node initiates the route discovery process only if the source needs to forward data packets to a destination, and the routing table of the source node do not have valid routes from the source to destination. So the source node first broadcasts route request (RREQ) packets to its neighbors. There are two different situations when a node receives a RREQ packet: i) this node sends a route reply (RREP) if it is the destination or it knows the route(s) to the destination; ii) the receiving node establishes a reverse route to the source if the routing table of this node does not have a routing entry for the destination. \par

After RREQ packets arrive at the destination, destination node unicasts a RREP packet to the source node from the selected reverse path. The route discovery process finishes when the source node receives the RREP message, and then data packets begin to be forwarded to the destination by the source node along the direction of established forwarding route.\par

The route maintenance procedure is operated by nodes in two different ways. One situation is that a node broadcasts hello messages to its neighbors at regular time intervals so that the node can maintain connectivity with its neighbors. Another situation is that the procedure aims to increase the successful data transmission ratio through the local repair mechanism \cite{Gulliver2005proceeding}. \par
\subsection{Overview of Software-defined Networking}
\label{sect:sys}

Software defined networking is an emerging network architecture where network control is decoupled from forwarding plane, and it can be directly programmable \cite{Xia2015ieeecst}. Because of the decoupling mechanism of the control and forwarding plane, network nodes only act as forwarding devices. Meanwhile, network control logic is moved into a logic control layer or a networking operating system \cite{Kreutz2015proceeding}.\par

\begin{figure}[tp]
\centering
\includegraphics[width=0.46\textwidth]{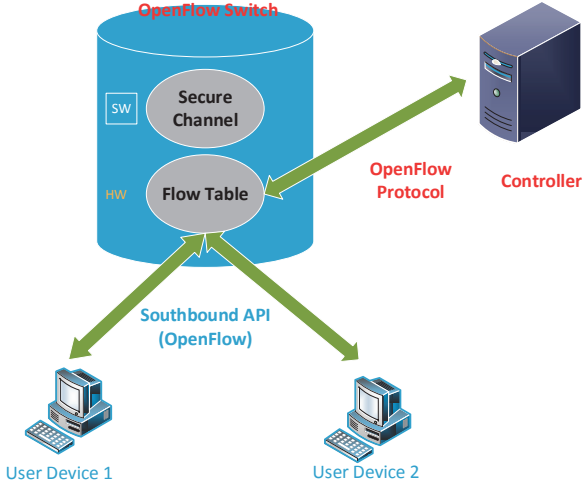}
\caption{OpenFlow switch and controller interaction using OpenFlow protocol.}
\label{fig:OpenFlow}
\end{figure}

There are many protocol standards on the use of SDN in real applications. One of the most famous protocol standards is called OpenFlow \cite{Kreutz2015proceeding}. OpenFlow is a widely used protocol that introduces the SDN concepts to implement in hardware and software. An prominent characteristic of OpenFlow is that the existing hardware can be utilized in SDN so as to design new protocols and to verify their feasibility \cite{Kreutz2015proceeding}. Figure \ref{fig:OpenFlow} shows the OpenFlow switch and controller.\par

The information interaction between OpenFlow switches and controller(s) supports three kinds of messages: controller-to-switch, asynchronous, and symmetric \cite{OpenFlowwhotepaper}. The most important message in control-to-switch is the \emph{OFPTFLOWMOD} \cite{OpenFlowwhotepaper}, which is used to modify the flow table in the OpenFlow switches. \emph{OFPTPACKETIN} \cite{OpenFlowwhotepaper} is the most important message in asynchronous, and this message enables the OpenFlow switches to send packets to controller only if the packets can not be processed by the switches. The most common message of the symmetric is named \emph{OFPTHELLO} \cite{OpenFlowwhotepaper}. It is used to build a connection between the OpenFlow switches and controller(s).
\section{AODV Protocol with Trust Based Mechanism (TAODV)}
 In this section, we describe our trust model. We assume that each node in our TAODV broadcasts packets to its neighbors periodically, and the neighbors receive the packets correctly. However, if a node broadcasts multiple packets at the same time, its neighbors only can receive a part of the packets because of some unexpected causes (such as heavy traffic) and malicious attacks (such as black-hole attack). We use a novel concept \emph{forwarding ratio} and node trust calculation process \cite {Li2010IET} to evaluate our node trust value.\par
\subsection{Node Trust Calculation Process}
\label{subsect:spectrum_sensing_model}
\emph{Definition 1(Forwarding ratio)}: Forwarding ratio is the number of packets received correctly divided by the number of packets forwarded. For example, we assume that a node $a$ sends 120 packets to its neighbor node $b$, and node $b$ only receives 100 packets because of the packet loss. Meanwhile, node $b$ only can forward 80 packets because of its transceiver capability, so the forwarding ratio of node $b$ to node $a$ is 0.8. The forwarding ratio $R_{ab}(t)$  of node $a$ to node $b$ can be defined by the following formula
\begin{equation}\label{eq:ctrPrefetching}
R_{ab}(t)=\frac{C_{ab}(t)}{T_{ab}(t)} \quad t\leq {W}
\end{equation}
where $C_{ab}(t)$ represents the number of the packets that a node can correctly forward to its neighbors. $T_{ab}(t)$ denotes the total number of packets that the node received before time $t$, where $W$ represents the width of the recent time window.\par

In TAODV, the packets can be divided into two groups: control packets (RREQ, RREP and RRER) and data packets. The control packets (RREQ and RREP) determine the data transfer path, and forwarding ratio of control packets is an important factor to determine node trust value. The node trust computation is shown below
\begin{equation}
\label{eq:defObjective}
N_{ab}(t)=\omega_1CR_{ab}(t)+\omega_2DR_{ab}(t)
\end{equation}
where $CR_{ab}(t)$ represents the control packet forwarding ratio and $DR_{ab}(t)$ represents the data packet forwarding ratio. $N_{ab}(t)$ denotes the trust value of receiving node $b$ for forwarding node $a$. ${\omega_1}$ and ${\omega_2}$ are two weighted factors (${\omega_1}, {\omega_2}\geq0$, and ${\omega_1+\omega_2=1}$) that determine which forwarding ratio ($CR_{ab}(t)$ and $DR_{ab}(t)$) is more important in the node trust calculating process. Particularly, we assume ${\omega_1=1}$ and ${\omega_2=0}$, which means the control packet forwarding ratio decides the overall node trust value.
\subsection{Path Trust Calculation Process}
\label{subsect:system.attack}
In the route discovery process, when a control packet such as RREQ arrives at a destination node, the routing path from source to destination is computed according to the node trust defined by the section 3.1. According to the axiom \cite{Sun12}, concatenation propagation of trust does not increase trust, the reverse and forwarding path trust value should not be more than the trust value of intermediate nodes. Meanwhile, since the control packet is a crucial factor to determine the node trust value, we add a new field called PacketTrust (PT) into the RREQ and RREP packet format and denote by $PT_{rreq}$ and $PT_{rrep}$. Specifically, we set the initial value of PacketTrust to 1. At time $t$, the trust value of a reverse path $P$ is denoted by $T_P(t)$ and given by the following formula
\begin{equation}
T_p(t)=N_{ab}(t)\times{PT^{rreq}_a}
\end{equation}
where $PT^{rreq}_{a}$ means the trust value of PacketTrust field in RREQ packet when a RREQ packet leaves node $a$.\par
\begin{equation}
T_p{'}(t)=N_{ba}(t)\times{PT^{rrep}_b}
\end{equation}
where $T_p{'}(t)$ denotes the trust value of a forwarding path. $PT^{rrep}_{b}$ means the trust value of PacketTrust field in RREP packet when a RREP packet leaves node $b$.
\subsection{The Objective Function of TAODV}
\label{subsect:system.trust}
 In our TAODV mechanism, there are two main factors influencing the whole network performance. One factor is hop count, and another is path trust value. Our goal is to evaluate the network performance in three different scenarios: the first one we only consider the path trust factor, the second one we consider both the hop count factor and path trust factor, the third one we only consider the hop count factor. So the objective function of our proposed TAODV protocol is shown in below: \par
\begin{equation}\label{eq:ctrBackhaulBS}
F(x)=\alpha x_1+\beta x_2
\end{equation}
where $F(x)$ denotes the network performance of VANETs using TAODV protocol. ${x_1}$ denotes the path trust value when control packets arrives at nodes, and ${x_2}$ denotes the hop count of control packets. ${x_1}$ and ${x_2}$ are two influence factors that determine the network performance when using TAODV protocol. ${\alpha}$ and ${\beta}$ (${\alpha,\beta\geq0}$) are two weighted factors.\par

From this equation, we make three assumptions that help us to analyze the TAODV mechanism:\par
1) When ${\alpha\gg\beta}$, we assume that the network performance is mainly decided by the path trust value ${x_1}$. \par
2) When ${\alpha\approx\beta}$, we assume that the network performance is decided by the path trust value  ${x_1}$ and hop count ${x_2}$. \par
3) When ${\alpha\ll\beta}$, we assume that the network performance is decided by the hop count. This scenario is the same as the original ad-hoc networks using AODV protocol. \par
\subsection{Route Discovery Process of TAODV}
\label{sect:framework}
The traditional AODV protocol aims to select a minimum hop count path to transfer the data packets. By contrast, in our TAODV protocol, we propose a trust based RREQ (T-RREQ) packet format, which contains the following fields:\par
\begin{flushleft}
(\emph{RREQID}, \emph{HopCount}, \emph{SourceAddr}, \emph{SourceSeq}, \emph{DestAddr},
\emph{DestSeq}, \emph{PacketTrust})
\end{flushleft}\par

As mentioned before, we design a new field named PacketTrust (PT), and add it into a RREQ packet. It is initialized to 1 and varies during the packet transmission process.\par
\begin{figure}[tp]
\centering
\includegraphics[width=0.46\textwidth]{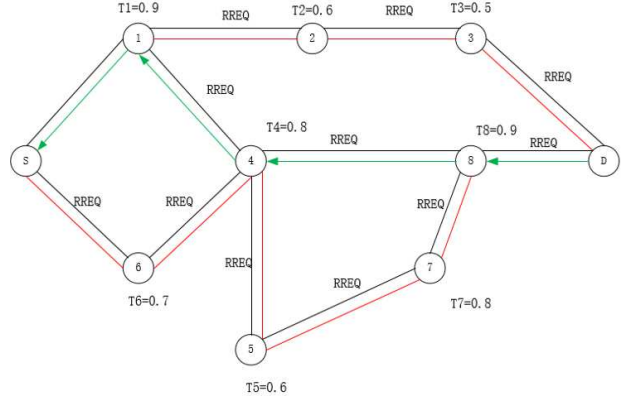}
\caption{An example of the TAODV calculation process.}
\label{fig4}
\end{figure}
In our TAODV protocol, when a node receives a T-RREQ packet from its neighbor, first this node checks the $RREQID$ of this T-RREQ. If the node has already received the same ID before, this T-RREQ is dropped by this node. On the contrary, if the $RREQID$ is new, the routing table of the node checks the sequence number in the RREQ packet, if the sequence number is fresh, the reverse path is established by the node and it updates its own routing table in which calculates the path trust value, if the sequence number is old, the node discards this RREQ packet. Meanwhile, when the node receives the T-RREQ message, it checks its routing table to determine whether this node is destination or have a fresh route to destination. If so, it updates its routing table and responds a T-RREP packet back to the source. If not, the node continues to broadcast this T-RREQ packet to its neighbors. If the node receives the different T-RREQ packets simultaneously, the node chooses a best path in the routing table with better path trust value. In other words, if the new path trust value is better than the previous one, the node updates the routing table immediately.\par
%%%%%%%%%%%%%%%%%%%%%%%%%%%%%%%%%%
%%%%%%%%%%%%%%%%%%%%%%%%%%%%%%%%%%%
Figure 2 shows an example of reverse path establishment process of TAODV. We assume that the source node need to initiate the route discovery process. The source first broadcasts T-RREQ packets to its neighbor node 1 and node 6. Meanwhile, the PacketTrust field in the T-RREQ is set to 1. The T-RREQ packets arrive at node 1 and node 6, the path trust is calculated in (4.3). The path trust value from source to node 1 is $T_{s1}=0.9\times{1}=0.9$. The path trust from source to node 6 is $T_{s6}=0.7\times{1}=0.7$. When node 1 and node 6 receive the T-RREQ packets, the value of PacketTrust field of the T-RREQ packets changes to 0.9. Node 4 receives two T-RREQ packets from node 6 and node 1. The routing table of node 4 compares the path trust value. Here the path trust $T_{64}=0.7\times{0.8}=0.56$ and $T_{14}=0.9\times{0.8}=0.72$. So node 4 discards the T-RREQ packet from node 6 because the path trust value from node 6 to node 1 is smaller than the path trust value from node 1 to node 4. After the path selection, node 4 sets up the reverse paths to the source. Similarly, the final reverse path is from destination, via node 7, node 4, and node 1 to the source. \par

When receiving a T-RREQ, the destination node replies T-RREP back to the source node via the intermediate nodes. Meanwhile, the forwarding paths are established when T-RREP packets pass through the switch nodes. The format of a T-RREP packet contains the following fields:\par
\begin{flushleft}
(\emph{HopCount}, \emph{SourceAddr}, \emph{SourceSeq}, \emph{DestAddr}, \emph{DestSeq},
\emph{PacketTrust}, \emph{lifetime})
\end{flushleft}\par
After receiving the RREP packet, the source sends the data packets following the forwarding path that established before to the destination node.
\section{Software-defined Vehicular ad-hoc Networks based on TAODV}
\label{sect:weighted_consensus}
In this section, we present a novel architecture SD-TAODV for data transmissions based on SDN. In the traditional AODV protocol, flow change of transmission packets due to the high node mobility occurs frequently \cite{Jo2016proceeding,WTY14}. Control logic and forwarding logic are all located on VANET nodes. By contrast, in our proposed SD-TAODV system, we move the control logic of VANETs from forwarding plane to a control plane in order to improve the network performance.
\subsection{Framework Description}
The framework of SD-TAODV is similar with the traditional SDN architecture. We divide the structure of SD-TAODV into three layers: (1) data forwarding plane operates the TAODV protocol and nodes in the plane supporting OpenFlow protocol; (2) NOS (controller) layer aims to manage the network topology and establishes the data transfer path for the data transmission; (3) application layer controls the forwarding rules, routing tables, and routing protocols. The whole SD-TAODV mechanism virtualizes the VANETs and provides the services for the application layer through the OpenFlow interfaces. \par
%So SD-TAODV enables the network to manage resources more flexible.

For the original OpenFlow structure as shown in Figure 1, no matter switches or controller, they are all fixed. However, the TAODV topology always keep changing because of the node mobility, so the architecture of SD-TAODV should be different from the traditional SDN.\par
\begin{figure}[tp]
\centering
\includegraphics[width=0.4\textwidth]{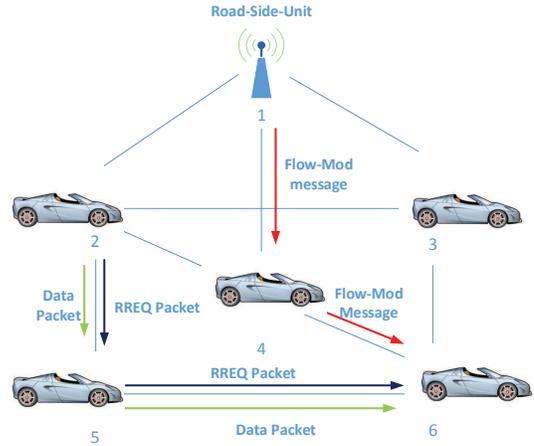}
\caption{An example vehicular ad-hoc network using SD-TAODV mechanism.}
\label{fig6}
\end{figure}

Briefly, if a switch node receives a TAODV control packet (T-RREQ or T-RREP), it sends the packet to the controller to handle. If a switch node receives a data packet, it forwards the packet to its neighbour node(s). The centralized control mechanism in SD-TAODV  manages the whole network in the control node. So the control node first needs to know the whole network topology.\par

\subsection{Network Topology Discovery}
The method of discovering the network topology is that the control node sends topology request messages to its neighbors. The topology request message includes the following fields:
\begin{flushleft}
(\emph{PacketID}, \emph{ControllerAddr}, \emph{NodeTrustList}, \emph{TopologyList})
\end{flushleft}\par
The ${NodeTrustList}$ is used to record the trust value of nodes when the message has passed by a node. When any one of nodes in the network topology receives the topology request message, the ${TopologyList}$ adds this node information into the ${TopologyList}$ field. Specifically, if a node receives the same ${PacketID}$ request message, the node sends back this packet to the controller immediately. Similarly, one of the nodes in the network topology sends the massage back to the controller if all of its neighbor nodes have already in the ${TopologyList}$ field. When the control node receives all the responses, the controller can establish the whole network topology or update the exciting topology. Figure 3 shows an example of the SD-TAODV network topology.  Road-side-unit 1 acts as a controller in the network topology. The method  to get the whole network topology is shown below:\par
1) Road-side-unit 1 initializes the topology request messages (e.g, the ID is 1357), and adds them into the ${TopologyList}$. Then road-side-unit 1 sends the request messages to its neighbors, node 2 and node 3.\par
2) Node 2 receives the request packet and adds itself into the ${TopologyList}$. Meanwhile, node 2 puts its own trust value into the ${NodeTrustList}$, and node 2 broadcasts the message to its neighbors. Due to the ${TopologyList}$ already has the information of node 1, node 2 only forwards the topology request message to nodes 3, 4 and 5. Similarly, node 3 only sends the request packet to its neighbor node 6. Meanwhile, node 3 sends the message back to node 2 because the ${PacketID}$ is the same. Node 4 receives the request message from node 2, and node 4 puts itself into the ${TopologyList}$. Since node 2 already exists in the ${TopologyList}$, node 4 only transfers the message to node 6. Analogously, node 5 also sends the request message to node 6.\par
3) Node 6 receives the request messages from nodes 3, 4 and 5. First node 6 puts its routing information into the ${TopologyList}$, and adds trust value into the ${NodeTrustList}$. Secondly, since the ${PacketID}$ of three packets is the same, node 6 sends back these three messages back to controller 1 according to the ${ControllerAddr}$.\par
4) When receiving all the responses from other nodes, road-side-unit 1 gets the whole network topology.

\subsection{The Working Process of the Controller}
After getting the network topology, the controller can control and manage the whole network. The information interaction between the controller and the OpenFlow switches includes the OpenFlow messages such as ${OFPTFlowMod}$ and ${OFPTPacketIN}$. The interfaces between the controller and OpenFlow switches are similar to the traditional southbound API. \par

When receiving an OpenFlow message from a forwarding node, the controller determines the type of the message. If the message is ${OFPTHello}$, the controller responds the message and builds a connection between the node and the controller. If the message is the ${OFPTPacketIN}$, the control node resolves the message and gets the message information, which includes the details of the T-RREQ and T-RREP packet. As we described before, when a T-RREQ or T-RREP packet arrives at the control node, the controller gets the value in ${PacketTrust}$ field, and calculates the path trust value. After finishing the packets handling, the controller sends an ${OFPTFlowMod}$ message to the forwarding node. \par
\subsection{The Working Process of the Forwarding Node}
The forwarding nodes in SD-TAODV are used to transfer the data packets and control packets. The interfaces between the control node and switches are the southbound API, which supports ${OFPTHello}$, ${OFPTFlowMod}$ and \\
\noindent ${OFPTPacketIn}$ messages.\par

In SD-TAODV, the forwarding nodes send the ${OFPTHello}$ messages to the control node periodically. If any node in the network topology receives the response from the control node, this forwarding node builds a connection with the control node. If a forwarding node receives a control packet such as T-RREQ packet from its neighbors, the T-RREQ packet first matches the flow table (we assume that T-RREQ and T-RREP cannot match the flow table). Otherwise, a forwarding node sends this packet to the controller in order to request a new flow table with the ${OFPTPacketIn}$ message. After resolving the packet, the control node responds an ${OFPTFlowMod}$ message back to the node and modifies the flow table, and  executes the action set in the flow table to handle this packet.\par
\section{Simulation Results and Discussions}
\label{sect:simulation}
In this section, we describe our simulation setup, configurations, and simulation results. OPNET is used as the simulator. In SD-TAODV simulations, we consider two different scenarios in the route discovery process: the first one only considers the path trust value factor, and the second one both considers the hop count factor and path trust value. Moreover, in our simulations, we assume that the nodes in our TAODV are all SDN-enabled. In addition, we assume that our SD-TAODV network includes two different types of nodes: i) normal nodes, which the data packets are normally forwarded by those nodes; ii) malicious nodes, which randomly drop the data packets when they receive the packets. The number of the malicious nodes is much smaller compared with the number of normal nodes. \par
\subsection{Simulation Setup}
Our simulation model is built on the OPNET Modeler and our model spans the area of ${5\times{5}{km}^2}$. We consider three different situations described in Section 3. In our simulation, the simulation parameters are as follows:\par
1) The simulation time is 15 mins.\par
2) The node density of our SD-TAODV model is 25 nodes.\par
3) The physical layer and MAC layer support IEEE 802.11.\par
There are three metrics evaluated in our simulations:\par
1) ${\emph{Average end-to-end delay}}$: the average end-to-end delay is the time calculated by the data packets to be transferred across the whole network from the source to the destination. It includes buffer delays during the route discovery process, queuing delays at interface queues, retransmission delays at MAC layers, and the propagation time from the source to the destination \cite{ZYL10,Li2010IET}.\par
2) ${\emph{Network throughput}}$: the throughput is the total size of packets received by the destination node at every second. The network throughput is an important factor to evaluate the network performance.\par
3) ${\emph{Total messages sent}}$: the total messages sent are the number of the routing messages sent in the entire network.\par

Firstly, the average end-to-end delay of the proposed SD-TAODV scheme is evaluated through the Figure 4 and 5. From these two figures, we can see that the end-to-end delay of SD-TAODV is higher than that of the traditional AODV with different data rates (1 Mbps, 2 Mbps, 5.5 Mbps and 11 Mbps). The reason why the end-to-end delay increasing as the data rates grow is that the quality of channels becomes more and more bad as the node velocities grow, so the risk probability of packet loss in the channels increases. As the data rates grow, the packets are easier to be dropped in the channels, so the end-to-end increasing as the data rates grow. There are two reasons for the higher end-to-end delay of SD-TAODV: i) in the route discover process, the SD-TAODV nodes always select fresh and higher path trust value routes to establish reverse and forwarding paths in order to transfer the data packets to the destination. The best trust value route reduces the risk probability of route breakdown because of the drop. However, the new routes may have more hop counts to the destination than the traditional AODV. The data packets need to spend more time to be transferred in the new routes; and ii) for the SD-TAODV scheme, the nodes first need to build connection with the control node. When TAODV packets (T-RREQ, T-RREP) arrive, nodes send the control packets to the controller to handle. This process also need to spend some time. So comparing with the traditional AODV network, the end-to-end delay of SD-TAODV is higher.\par
Figures 6 and 7 depict the throughput comparison of SD-TAODV and traditional AODV in different data rates. In Figure 6, we only consider the trust value factor. In Figure 7, we consider the trust value and hop count. Through these two figures, we can conclude that the performance of the proposed SD-TAODV mechanism is better than the traditional AODV protocol. These two scenarios indicates that the network performance of SD-TAODV is better than the traditional AODV network. As the data rates grow, we can find that the throughput of both SD-TAODV and original AODV all increases. This is because  more data packets can be received as the data rates grow. The reason for the better performance of SD-AOTDV is that the best trust value path is selected by the SD-AOTDV system, which means that the selected path between two nodes reduces the risk of packet loss and the quality of links are better than the traditional AODV. In other words, as the data packets are transferred on the secure paths, the possibilities of packets loss are lower than the traditional AODV. So the SD-TAODV scheme has performance improvement in terms of network throughput compared to traditional AODV. \par

\subsection{Evaluation}
\begin{figure}[tp]
\centering{
\includegraphics[width=0.5\textwidth]{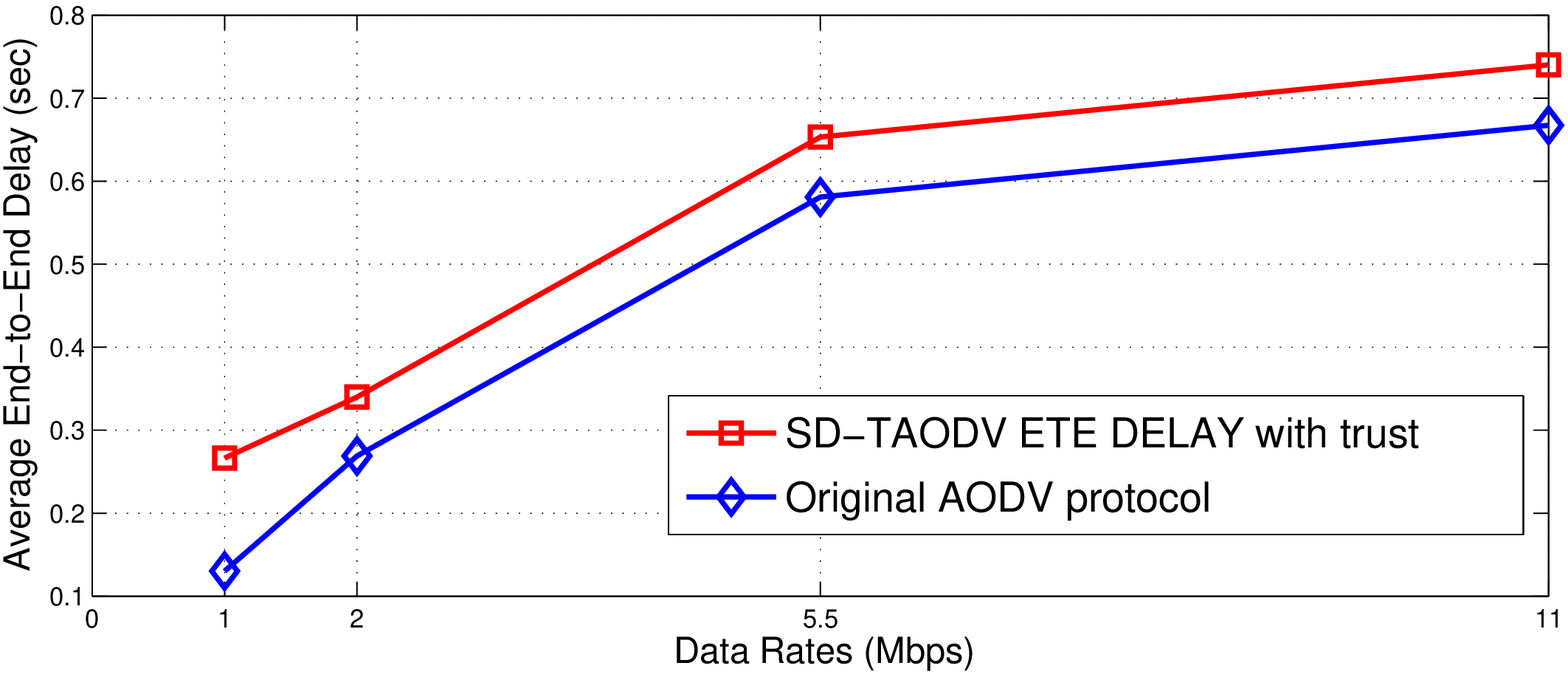}
\caption{Average ETE delay of SD-TAODV with trust value in different data rates.}
\label{pic:existing_ss_malicious}}
%\end{figure}
%\begin{figure}[tp]
\centering{
\includegraphics[width=0.5\textwidth]{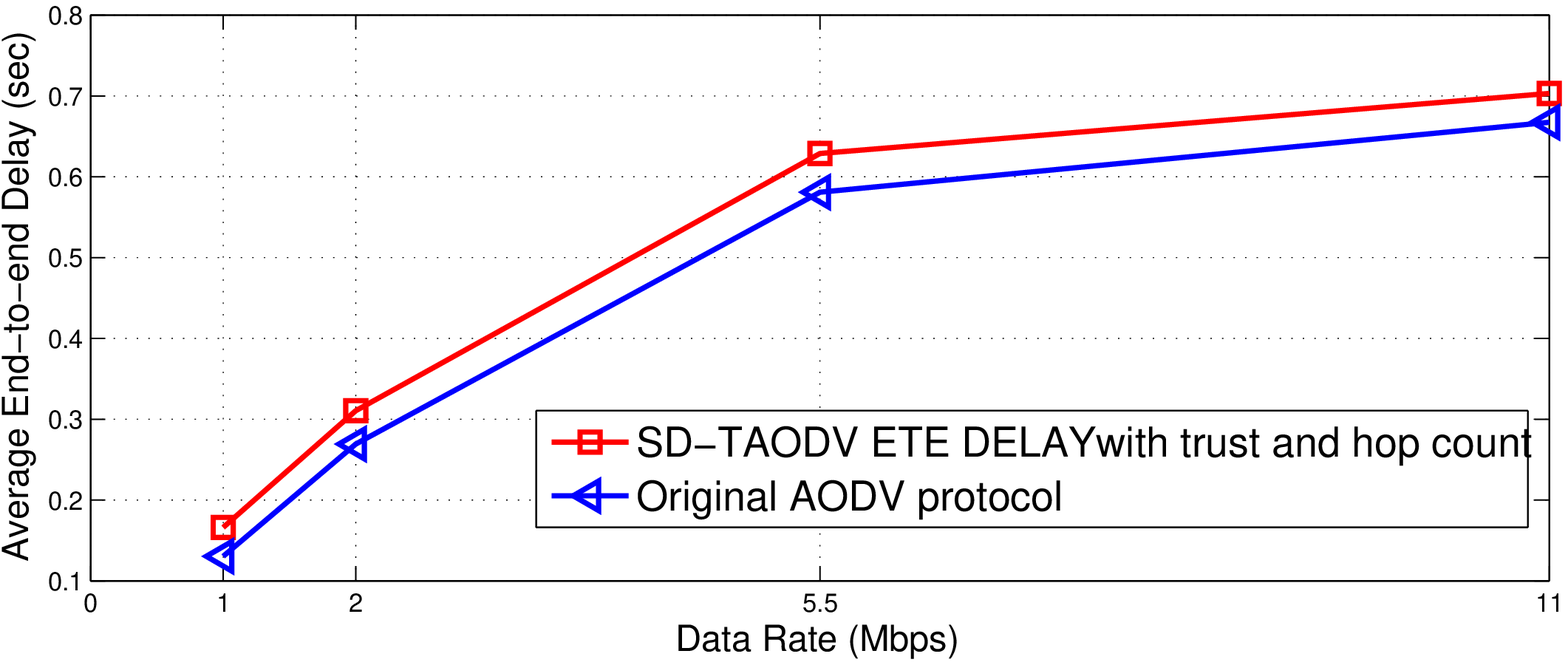}
\caption{Average ETE delay of SD-TAODV with trust value and hop count in different data rates.}
\label{pic:proposed_ss_malicious}}
%\end{figure}
%
\end{figure}

In addition, we also evaluate the SD-TAODV network performance in different numbers of VANET nodes. In Figure 8, we can see that the throughput of SD-TAODV and original AODV all decreases as the number of nodes grows. We assume that the number of malicious nodes increases as the number of nodes grows. The malicious nodes also have big impact on the throughput of SD-TAODV network \cite{Wei2014ieeetvt}. The network throughput decreases significantly, as shown in Fig. 8. Although the network throughput decreases as the number of nodes grows, the SD-TAODV network throughput still better than the original scheme. Because network nodes update their neighbors' information periodically, the control node of SD-TAODV can respond faster to the topology change. As the network nodes leave or join the network, the control node detects the topology change and sends the control messages to these new nodes to maintain the data transfer path. So our proposed scheme has performance improvement than the traditional AODV.\par

Finally, we compare the total message overhead sent in different number of nodes. Figure 9 depicts how much message overhead sent by the SD-TAODV mechanism compared with the traditional AODV protocol. Through Figure 9, we can conclude that the message overhead of the SD-TAODV is higher than the original AODV. This is because the nodes in SD-TAODV network need to send extraneous messages such as ${OFPTHello}$ and ${OFPTPacketIn}$ to the control node. In Figure 9, since more nodes join the network as the number of nodes increases, the message overhead grows simultaneously in both AODV and SD-TAODV. \par
\begin{figure}[tp]
\centering
\includegraphics[width=0.5\textwidth]{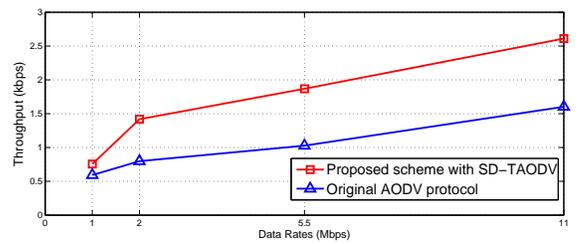}
\caption{Average throughput comparison of SD-TAODV with the trust value.}
\label{fig11}
\end{figure}

\begin{figure}[tp]
\centering
\includegraphics[width=0.5\textwidth]{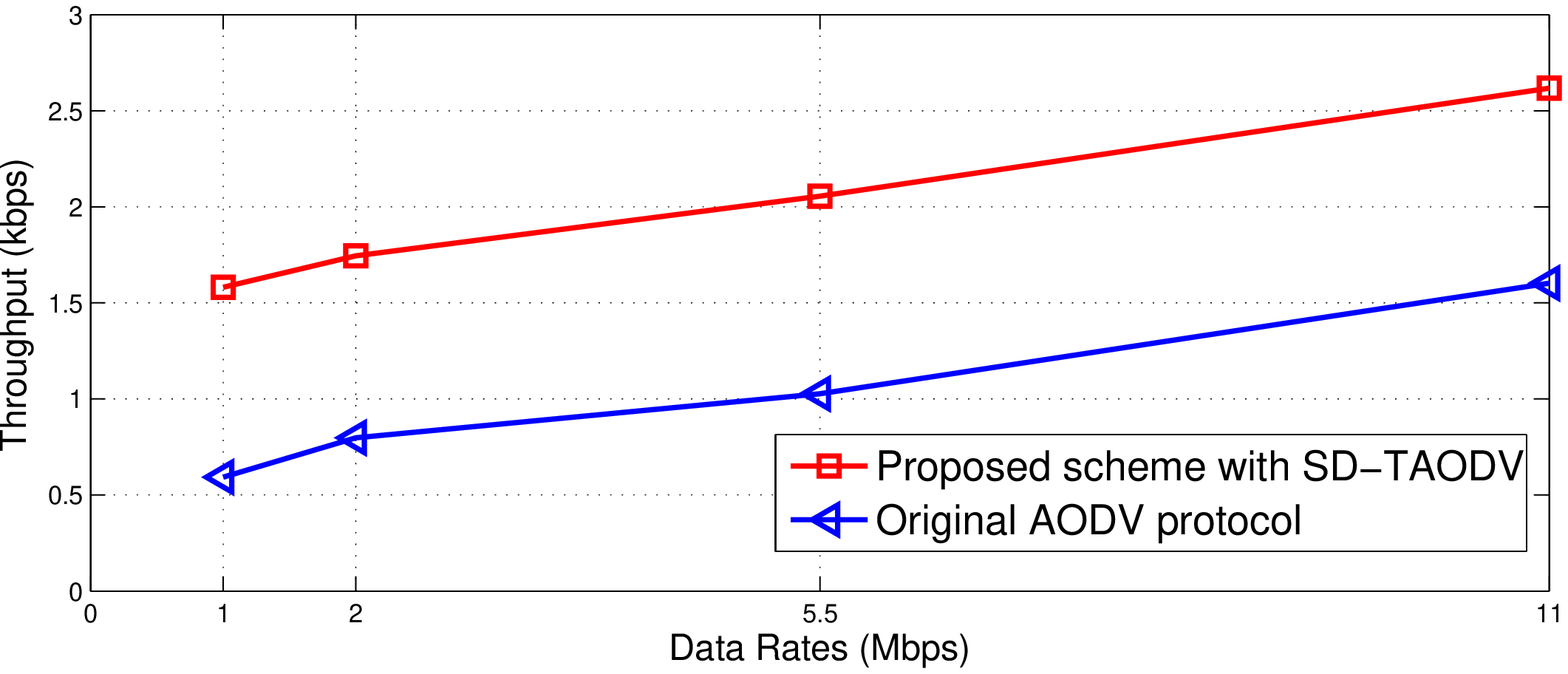}
\caption{Average throughput comparison of SD-TAODV with the trust value and hop count.}
\label{fig12}
\end{figure}

\begin{figure}[tp]
\centering
\includegraphics[width=0.5\textwidth]{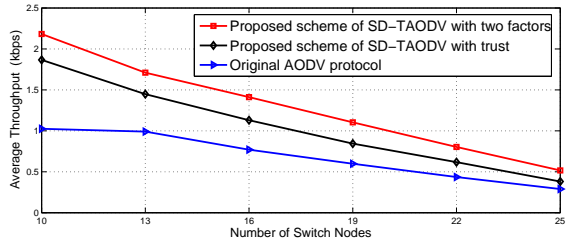}
\caption{Average throughput comparison with different numbers of nodes.}
\label{fig15}
\end{figure}

\begin{figure}[tp]
\centering
\includegraphics[width=0.5\textwidth]{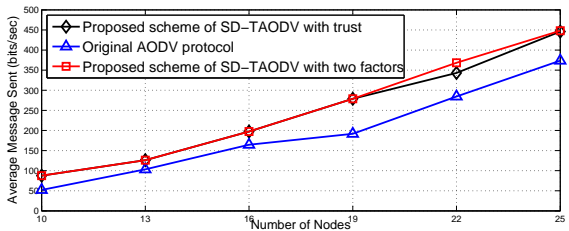}
\caption{Total message overhead comparison with different numbers of nodes.}
\label{fig16}
\end{figure}

\section{Conclusions and Future Work}
\label{sect:dso}

In this paper, we have presented a novel framework of software-defined VANETs with trust management. We designed a routing protocol named software-defined trust based ad hoc on-demand distance vector routing (SD-TAODV). In SD-TAODV, the route discovery and the route maintenance process are moved into a controller, and the reverse and forwarding paths are chosen by the controller. Simulation results were presented to show the effectiveness of the proposed software-defined VANETs with trust management. We compared our SD-TAODV protocol with the traditional AODV protocol in terms of end-to-end delay, throughput and message overhead. Although the end-to-end delay of SD-TAODV is higher than AODV, the network throughput performance improves significantly in SD-TAODV. In the future, we will study to reduce the end-to-end delay of our proposed framework. \par

\section*{Acknowledgment}

This work was supported in part by the Natural Sciences and Engineering Research Council of Canada (NSERC) DIVA Network.

\balance
\bibliographystyle{unsrt}
\bibliography{references}

\begin{thebibliography}{10}

\bibitem{Silva13}
A.~Silva Fabr{\'i}cio, Celes Clayson, and Azzedine Boukerche.
\newblock Filling the gaps of vehicular mobility traces.
\newblock In {\em Proc. 18th ACM Int'l Conference on Modeling, Analysis and
  Simulation of Wireless and Mobile Systems (MSWiM'15)}, pages 47--54, New
  York, NY, USA, 2015. ACM.

\bibitem{Zhuang13}
Abboud Khadige and W.~Zhuang.
\newblock Impact of node mobility on single-hop cluster overlap in vehicular ad
  hoc networks.
\newblock In {\em Proc. 17th ACM Int'l Conference on Modeling, Analysis and
  Simulation of Wireless and Mobile Systems (MSWiM'14)}, pages 65--72, 2014.

\bibitem{Nidhal13}
M.~N. Mejri and Jalel Ben-Othman.
\newblock Entropy as a new metric for denial of service attack detection in
  vehicular ad-hoc networks.
\newblock In {\em Proc. 17th ACM Int'l Conference on Modeling, Analysis and
  Simulation of Wireless and Mobile Systems (MSWiM'14)}, pages 73--79, 2014.

\bibitem{GDS13}
Agata Grzybek, Gr{\'e}goire Danoy, Marcin Seredynski, and Pascal Bouvry.
\newblock Evaluation of dynamic communities in large-scale vehicular networks.
\newblock In {\em Proc. Third ACM Int'l Symp. Design and Analysis of
  Intelligent Vehicular Networks and Applications (DIVANet'13)}, pages 93--100,
  2013.

\bibitem{MYL04}
L.~Ma, F.~Yu, V.~C.~M. Leung, and T.~Randhawa.
\newblock A new method to support {UMTS/WLAN} vertical handover using {SCTP}.
\newblock {\em IEEE Wireless Commun.}, 11(4):44--51, Aug. 2004.

\bibitem{YL01}
Fei Yu and V.~C.~M. Leung.
\newblock Mobility-based predictive call admission control and bandwidth
  reservation in wireless cellular networks.
\newblock In {\em Proc. IEEE INFOCOM'01}, Anchorage, AK, Apr. 2001.

\bibitem{LYH10}
Zhiqiang Li, F.~Richard Yu, and Minyi Huang.
\newblock A distributed consensus-based cooperative spectrum-sensing scheme in
  cognitive radios.
\newblock {\em IEEE Trans.\ Veh.\ Tech.}, 59(1):383--393, 2010.

\bibitem{YK07}
F.~Yu and V.~Krishnamurthy.
\newblock Optimal joint session admission control in integrated {WLAN} and
  {CDMA} cellular networks with vertical handoff.
\newblock {\em IEEE Trans. Mobile Computing}, 6(1):126--139, Jan. 2007.

\bibitem{XYJL12}
R.~Xie, F.~R. Yu, Hong Ji, and Yi~Li.
\newblock Energy-efficient resource allocation for heterogeneous cognitive
  radio networks with femtocells.
\newblock {\em IEEE Trans. Wireless Commun.}, 11(11):3910 --3920, Nov. 2012.

\bibitem{ATV12}
A.~Attar, H.~Tang, A.V. Vasilakos, F.~R. Yu, and V.C.M. Leung.
\newblock A survey of security challenges in cognitive radio networks:
  Solutions and future research directions.
\newblock {\em Proceedings of the IEEE}, 100(12):3172--3186, 2012.

\bibitem{LY15}
C.~Liang and F.~R. Yu.
\newblock Wireless network virtualization: A survey, some research issues and
  challenges.
\newblock {\em IEEE Commun. Surveys Tutorials}, 17(1):358--380, Firstquarter
  2015.

\bibitem{WYS10}
Yifei Wei, F.~R. Yu, and Mei Song.
\newblock Distributed optimal relay selection in wireless cooperative networks
  with finite-state {Markov} channels.
\newblock {\em IEEE Trans. Veh. Tech.}, 59(5):2149 --2158, June 2010.

\bibitem{GYJ10}
Quansheng Guan, F.~R. Yu, Shengming Jiang, and Gang Wei.
\newblock Prediction-based topology control and routing in cognitive radio
  mobile ad hoc networks.
\newblock {\em IEEE Trans. Veh. Tech.}, 59(9):4443 --4452, Nov. 2010.

\bibitem{BYC12}
S.~Bu, F.~R. Yu, Y.~Cai, and P.~Liu.
\newblock When the smart grid meets energy-efficient communications: Green
  wireless cellular networks powered by the smart grid.
\newblock {\em IEEE Trans. Wireless Commun.}, 11:3014--3024, Aug. 2012.

\bibitem{XYJ12}
R.~Xie, F.~R. Yu, and H.~Ji.
\newblock Dynamic resource allocation for heterogeneous services in cognitive
  radio networks with imperfect channel sensing.
\newblock {\em IEEE Trans. Veh. Tech.}, 61:770--780, Feb. 2012.

\bibitem{YTH09}
F.~R. Yu, H.~Tang, M.~Huang, Z.~Li, and P.~C. Mason.
\newblock Defense against spectrum sensing data falsification attacks in mobile
  ad hoc networks with cognitive radios.
\newblock In {\em Proc. IEEE Military Commun. Conf. (MILCOM)'09}, Oct. 2009.

\bibitem{YHT10}
F.~R. Yu, Minyi Huang, and H.~Tang.
\newblock Biologically inspired consensus-based spectrum sensing in mobile ad
  hoc networks with cognitive radios.
\newblock {\em IEEE Network}, 24(3):26 --30, May 2010.

\bibitem{LYJ10}
C.~Luo, F.~R. Yu, H.~Ji, and V.~C.~M. Leung.
\newblock Cross-layer design for {TCP} performance improvement in cognitive
  radio networks.
\newblock {\em IEEE Trans. Veh. Tech.}, 59(5):2485--2495, 2010.

\bibitem{YZX11}
F.~R. Yu, Peng Zhang, Weidong Xiao, and P.~Choudhury.
\newblock Communication systems for grid integration of renewable energy
  resources.
\newblock {\em IEEE Network}, 25(5):22 --29, Sept. 2011.

\bibitem{LYL09}
J.~Liu, F.~R. Yu, C.-H. Lung, and H.~Tang.
\newblock Optimal combined intrusion detection and biometric-based continuous
  authentication in high security mobile ad hoc networks.
\newblock {\em IEEE Trans. Wireless Commun.}, 8(2):806--815, 2009.

\bibitem{ZYN12_JSAC}
L.~Zhu, F.~R. Yu, B.~Ning, and T.~Tang.
\newblock Cross-layer handoff design in {MIMO}-enabled {WLANs} for
  communication-based train control ({CBTC}) systems.
\newblock {\em IEEE J. Sel. Areas Commun.}, 30(4):719--728, May 2012.

\bibitem{Tyagi2014proceeding}
P.~Tyagi and D.~Dembla.
\newblock Investigating the security threats in vehicular ad hoc networks
  ({VANETs}): Towards security engineering for safer on-road transportation.
\newblock In {\em Proc. IEEE Int'l Conf. Advances in Computing, Communications
  and Informatics (ICACCI)}, pages 2084--2090, New Delhi, 2014.

\bibitem{Wei13}
Z.~Wei, F.~Richard Yu, and A.~Boukeche.
\newblock Cooperative spectrum sensing with trust assistance for cognitive
  radio vehicular ad hoc networks.
\newblock In {\em Proc. Fifth ACM Int'l Symp. Design and Analysis of
  Intelligent Vehicular Networks and Applications (DIVANet'15)}, pages 27--33,
  2015.

\bibitem{Richard13}
Z.~Wei, F.~Richard Yu, and Azzedine Boukerche.
\newblock Trust based security enhancements for vehicular ad hoc networks.
\newblock In {\em Proc. Fourth ACM Int'l Symp. Design and Analysis of
  Intelligent Vehicular Networks and Applications (DIVANet'14)}, pages
  103--109, 2014.

\bibitem{Wang13}
Y.~Wang, F.~Richard Yu, M.~Huang, and T.~Chen.
\newblock Securing vehicular ad hoc networks with mean field game theory.
\newblock In {\em Proc. Third ACM Int'l Symp. Design and Analysis of
  Intelligent Vehicular Networks and Applications (DIVANet'13)}, pages 55--60,
  2013.

\bibitem{Boukerche13}
D.~Zheng, F.~Richard Yu, and A.~Boukerche.
\newblock Security and quality of service ({QoS}) co-design using game theory
  in cooperative wireless ad hoc networks.
\newblock In {\em Proc. Second ACM Int'l Symp. Design and Analysis of
  Intelligent Vehicular Networks and Applications (DIVANet'12)}, pages
  139--146, 2012.

\bibitem{HKH10}
Kuan~Lun Huang, Salil~S. Kanhere, and Wen Hu.
\newblock Are you contributing trustworthy data?: the case for a reputation
  system in participatory sensing.
\newblock In {\em Proc. 13th ACM Int'l Conference on Modeling, Analysis, and
  Simulation of Wireless and Mobile Systems (MSWIM'10)}, New York, NY, USA,
  2013.

\bibitem{RPB11}
Yonglin Ren, Richard~W.~N. Pazzi, and Azzedine Boukerche.
\newblock Outlier detection using naïve bayes in wireless ad hoc networks.
\newblock In {\em Proc. First ACM Int'l Symp. Design and Analysis of
  Intelligent Vehicular Networks and Applications (DIVANet'11)}, New York, NY,
  USA, 2011.

\bibitem{SKM12}
Daniel~Da Silva, Tracy~Ann Kosa, Steve Marsh, and Khalil El-Khatib.
\newblock Examining privacy in vehicular ad-hoc networks.
\newblock In {\em Proc. Second ACM Int'l Symp. Design and Analysis of
  Intelligent Vehicular Networks and Applications (DIVANet'12)}, New York, NY,
  USA, 2012.

\bibitem{NTH13}
Hasen Nicanfar, Peyman TalebiFard, Seyedali Hosseininezhad, Victor~C.M. Leung,
  and Mark Damm.
\newblock Security and privacy of electric vehicles in the smart grid context:
  problem and solution.
\newblock In {\em Proc. Third ACM Int'l Symp. Design and Analysis of
  Intelligent Vehicular Networks and Applications (DIVANet'13)}, New York, NY,
  USA, 2013.

\bibitem{Kreutz2015proceeding}
D.~Kreutz, M.V. Ramos, P.E. Verissimo, C.E. Rothenberg, S.~Azodolmolky, and
  S.~Uhlig.
\newblock Software-defined networking: A comprehensive survey.
\newblock {\em Proc. IEEE}, 103(1):14--76, January 2015.

\bibitem{YYG15}
Qiao Yan, F.~R. Yu, Qingxiang Gong, and Jianqiang Li.
\newblock Software-defined networking ({SDN}) and distributed denial of service
  ({DDoS}) attacks in cloud computing environments: A survey, some research
  issues, and challenges.
\newblock {\em IEEE Commun. Survey and Tutorials}, 18(1):602--622, 2016.

\bibitem{LYZ15}
C.~Liang, F.~R. Yu, and X.~Zhang.
\newblock Information-centric network function virtualization over {5G} mobile
  wireless networks.
\newblock {\em IEEE Network}, 29(3):68--74, May 2015.

\bibitem{LY15m}
C.~Liang and F.~R. Yu.
\newblock Wireless virtualization for next generation mobile cellular networks.
\newblock {\em IEEE Wireless Comm.}, 22(1):61--69, Feb. 2015.

\bibitem{CYL15}
Y.~Cai, F.~R. Yu, C.~Liang, B.~Sun, and Q.~Yan.
\newblock Software defined device-to-device {(D2D)} communications in virtual
  wireless networks with imperfect network state information {(NSI)}.
\newblock {\em IEEE Trans. Veh. Tech.}, 2015.
\newblock DOI:10.1109/TVT.2015.2483558.

\bibitem{CYY15}
Laizhong Cui, F.~R. Yu, and Qiao Yan.
\newblock When big data meets software-defined networking ({SDN}): {SDN} for
  big data and big data for {SDN}.
\newblock {\em IEEE Network}, 30(1):58--65, Jan. 2016.

\bibitem{Xia2015ieeecst}
W.~Xia, Y.~Wen, C.H. Foh, D.~Niyato, and H.~Xie.
\newblock A survey on software-defined networking.
\newblock {\em IEEE Communications Surveys and Tutorials}, 17(1):27--51, March
  2015.

\bibitem{Perkins2003framework}
C.~Perkins, E.~Belding-Royer, and S.~Das.
\newblock Ad hoc on-demand distance vector ({AODV}) routing.
\newblock Technical report, RFC 3561, July 2003.

\bibitem{Toulgoat13}
J.~Li, M.~Toulgoat, M.~Deziel, F.~Richard Yu, and S.~Perras.
\newblock Propagation modeling and mac-layer performance in {EM}-based
  underwater sensor networks.
\newblock In {\em Proc. Fourth ACM Int'l Symp. Design and Analysis of
  Intelligent Vehicular Networks and Applications (DIVANet'14)}, pages
  111--117, 2014.

\bibitem{Paul2011proceeding}
B.~Paul, M.~Lbrahim, and M.~A.~N. Bikas.
\newblock {VANET} routing protocols: Pros and cons.
\newblock {\em Int'l Journal of Computer Applications}, 20(3):28--34, April
  2011.

\bibitem{Gulliver2005proceeding}
Y.~Zhang and T.A. Gulliver.
\newblock Quality of service for ad hoc on-demand distance vector routing.
\newblock In {\em Proc. IEEE Int'l Conf. Wireless And Mobile Computing,
  Networking And Communications (WiMob)}, pages 192--196, Montreal, Canada,
  August 2005.

\bibitem{Cao2010proceeding}
Z.~Cao and G.~Lu.
\newblock {S-AODV}: Sink routing table over {AODV} routing protocol for
  6{L}o{WPAN}.
\newblock In {\em Proc. IEEE Int'l Conf. Networks Security Wireless
  Communications and Trusted Computing (NSWCTC)}, pages 340--343, Wuhan, China,
  April 2010.

\bibitem{OpenFlowwhotepaper}
Openflow switch specification.
\newblock Website, Dec. 2008.
\newblock http://archive.openflow.org/documents/openflow-spec-v0.8.9.pdf.

\bibitem{Li2010IET}
X.~Li, Z.~Jia, L.~Wang, and H.~Wang.
\newblock Trust-based on-demand multipath routing in mobile ad hoc networks.
\newblock {\em IET Information Security}, 4(4):212--232, December 2010.

\bibitem{Sun12}
Y.L. Sun, W.~Yu, Z.~Han, and K.J.~Ray Liu.
\newblock Information theoretic framework of trust modeling and ealuation for
  ad hoc networks.
\newblock {\em IEEE J. Selected Areas in Communications}, 24(2):212--232, 2006.

\bibitem{Jo2016proceeding}
M.Y. Jo and K.~Kim.
\newblock A research on the regional routing scheme based mobile agent for
  {SDN}.
\newblock In {\em Proc. IEEE Int'l Conf. Information Networking (ICOIN)}, pages
  211--213, Kota, Kinabalu, January 2016.

\bibitem{WTY14}
F.Yu, V.~W.~S. Wong, and Victor Leung.
\newblock Efficient {QoS} provisioning for adaptive multimedia in mobile
  communication networks by reinforcement learning.
\newblock {\em Mobile Networks and Applications}, 11(1):101--110, Feb. 2006.

\bibitem{ZYL10}
S.~Zhang, F.~R. Yu, and V.C.M. Leung.
\newblock Joint connection admission control and routing in {IEEE} 802.16-based
  mesh networks.
\newblock {\em IEEE Trans.\ Wireless Commun.}, 9(4):1370 --1379, Apr. 2010.

\bibitem{Wei2014ieeetvt}
Z.~Wei, H.~Tang, F.~Richard Yu, M.~Wang, and P.~Mason.
\newblock Security enhancements for mobile ad hoc networks with trust
  management using uncertain reasoning.
\newblock {\em IEEE Trans.\ Veh.\ Tech.}, 63(9):4647--4658, November 2014.

\end{thebibliography}
\end{document}